# Deep Learning Systems for Advanced Driving Assistance

**Francesco Rundo**
STMicroelectronics, ADG Power and Discretes R&D
francesco.rundo@st.com

## Abstract

Next generation cars embed intelligent assessment of car driving safety through innovative solutions often based on usage of artificial intelligence. The safety driving monitoring can be carried out using several methodologies widely treated in scientific literature. In this context, the author proposes an innovative approach that uses ad-hoc bio-sensing system suitable to reconstruct the physio-based attentional status of the car driver. To reconstruct the car driver physiological status, the author proposed the use of a bio-sensing probe consisting of a coupled LEDs at Near infra-Red (NiR) spectrum with a photodetector. This probe placed over the monitored subject allows to detect a physiological signal called PhotoPlethysmoGraphy (PPG). The PPG signal formation is regulated by the change in oxygenated and non-oxygenated hemoglobin concentration in the monitored subject's bloodstream which will be directly connected to cardiac activity in turn regulated by the Autonomic Nervous System (ANS) that characterizes the subject's attention level. This so designed car driver drowsiness monitoring will be combined with further driving safety assessment based on correlated intelligent driving scenario understanding.

## 1. Introduction

Physiological signals represent a relevant data source to be employed for the development of promising applications from cardiology to industrial and automotive fields [Tomohiko et al, 2015]. The increasing proliferation of non-invasive devices for collecting physiological parameters has led to explore various approaches to analyze physiological signals taking advantage of these new technologies instead of using invasive tools. In this context, PhotoPlethysmo-Graphic (PPG) signal has been proposed as a valid solution to analyze a subject's physiological status [Tomohiko *et al*, 2015]-[Vicente *et al*, 2011]. PPG is a convenient and simple physiological signal that provides information about the cardiac/neuro-physio activity of a subject [Rundo *et al*, 2018]. In the automotive industry, the increasing improvements in safety awareness systems have led to the development of ADAS (Advanced Driving Assistance Systems) architectures based on continuous monitoring of the driver's drowsiness by using physiological signals. Drowsiness refers to a physiological state characterized by the reduction of the level of consciousness and difficulty in maintaining the wakeful state [Rundo *et al*, 2018; Vinciguerra er al, 2019]. Cardiac activity (specifically, the heart rate) is regulated by the Autonomic Nervous System (ANS) through the action of the sympathetic/parasympathetic sub-systems [Rundo *et al*, 2018]. The action of the respective ANS sub-systems is related to the attentional state of the subject, which is reflected in the cardiac activity and, therefore, in the physiological heart-driven signals such as the PPG. Based on these assumptions, the study of cardiac activity and, therefore, of the correlated physiological signals, represents a great tool for monitoring drowsiness status of a subject as well as pathologies which may indirectly have an impact on the subject's guidance [Rundo *et al*, 2018, Vinciguerra er al, 2019]. Specifically, we focused on the use of the PPG signal for monitoring the subject's blood pressure and drowsiness level. Several studies have confirmed that a system for monitoring the risk of driving in an automotive environment can be obtained by monitoring the so-called "driver fatigue condition" which in turn is connected to a careful measurement of the level of physiological attention combined with the pressure level [Tomohiko et al, 2015; Rundo *et al*, 2018; Vicente *et al*, 2011; Vinciguerra er al, 2019]. The PPG based car driver drowsiness monitoring will be combined with further driving safety assessment systems. We propose ad-hoc delivered use-case: Detection and tracking of salient pedestrians embedded in the driving scenario.

## 2. The physio-based driver assessment system

As introduced, the main usefulness of the PPG signal consists in having a non-invasive car driver data sampling approach for monitoring the drowsiness of the subject. The pulsatile 'AC' component of the PPG signal is formed by a physiological impulse regulated by synchronized pulsing of the heartbeat. The remaining slowly changing component ('DC') contains information about the respiratory act, the thermoregulation activity and so on. With each contraction the heart exerts an adequate force for the right distension of arterioles and capillaries. If we have a device with a light emitter and detector embedded to the subject's skin, the pressure on the capillaries caused by the heartbeat can be detected at each pressure peak. More in detail, the heart pump activity causes in the subcutaneous capillary a pressure gradient that stretches the small vessel. In this physiological phase, if the arteries will be hit by a certain light source, some of this will be reflected (back-scattered) and then can be acquired by ad-hoc closed detector [Rundo et al, 2018, Conoci *et al*, 2018]. In this way the back-scattered light information will be transduced into an electrical signal (the final PPG signal) which can be easily processed by the downstream system. Further details in [Rundo et al, 2018, Conoci *et al*, 2018]. In Fig 1 we report an instance of the PPG signal formation process we have outlined in the previous description.

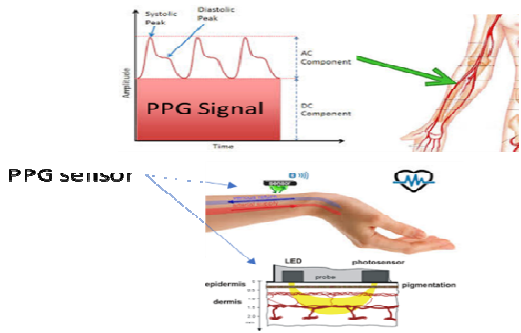

**Figure 1**. The PPG signal formation process with a detail of the sampling system

The proposed pipeline is equipped with a PPG sensing probe which samples data through a coupled LEDs - Silicon PhotoMultiplier (SiPM) device delivered by STMicroelectronics [Rundo *et al*, 2018; Conoci *et al*, 2018; Mazzillo *et al,* 2018]. The proposed PPG probes comprises an array device, called Silicon Photomultipliers(SiPMs) characterized by a total area of 4.0×4.5mm$^2$ and 4871 square microcells with 60μm pitch [Mazzillo *et al,* 2018]. As introduced, the PPG detector embeds also a light emitters [Rundo *et al*, 2018; Conoci *et al*, 2018; Mazzillo *et al,* 2018]. Ad-hoc electronic system to sample and buffer the generated PPG data will be implemented [Conoci *et al*, 2018; Mazzillo *et al,* 2018; Rundo *et al*, 2019b; Rundo *et al*, 2021]. In order to sample the car driver PPG signal, we placed such PPG sensor probes on the car steering wheel. To properly collect the physiological signal, the driver has to maintain only one hand on top of the embedded PPG sensor probes, to trigger the signal. Ad-hoc pre-processing pipeline has been designed to properly filter and process the collected raw PPG samples. To this end, a SPC58x MCUs series developed by STMicroelectronics boosted with a STA1295 Accordo5 MCU based system has been implemented for hosting the designed artificial intelligence algorithms [Mazzillo *et al,* 2018; Rundo *et al*, 2019b]. A schematic overview of the designed PPG-based drowsiness monitoring system is reported in Fig. 2.

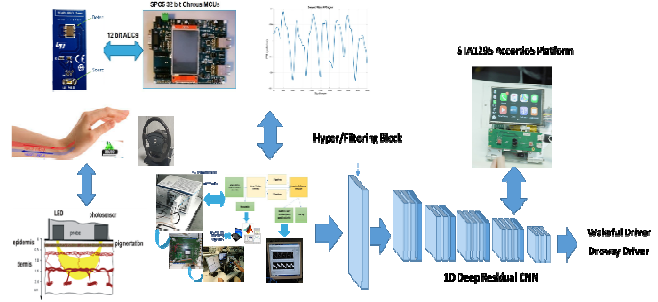

**Figure 2**. The PPG sampling hardware system

The so collected driver PPG waveforms (raw data) of the car driver will be collected in the STA1295 MCU system as per scheme reported in figure 2. Furthermore, considering the real scenario relating to driving on the road, it is plausible to hypothesize the presence of several type of noises due to the movement of the driver's hands or road-driven vibrations. To this end, both in terms of frequency filtering and signal stabilization, the author has proposed a bio-inspired pipeline that allows to obtain excellent results as widely documented in the work reported in [Trenta *et al*, 2019a; Rundo *et al*, 2019b; Rundo *et al*, 2021]. The PPG frequency range is 1-10 Hz.

To perform an effective filtering between 1-10 Hz, a further bio-inspired algorithm described in [Rundo *et al*, 2019b; Rundo *et al*, 2021] has been implemented to perform intelligent stabilization of the PPG signal. This algorithm has been named "Hyper-filtering Filter System" and it was inspired by the classical hyper-spectral approach [Trenta *et al*, 2019a; Rundo *et al*, 2019b; Rundo *et al*, 2021]. The hyper-spectral processing usually applied to 2D imaging is able to process visual data leveraging features from the filtering of the image across to the whole electromagnetic spectrum. With aims to apply the hyper-spectral theory to 1D signal, the author discovered the hyper-filtering approach [Trenta *et al*, 2019a; Rundo *et al*, 2019b; Rundo *et al*, 2021]. The information derived from hyper-filtering at different frequency ranges of the source PPG signal was analyzed to try to obtain information characterizing the "frequency spectrum of each signal sample" to be used to assess the driver's attention level (Drowsiness monitoring). Consequently, each layer of Hyper filtering has been divided into 11 sub-bands. Once the number of sub-bands has been established, a deep Reinforcement Learning [Trenta *et al*, 2019a; Rundo *et al*, 2019b; Rundo *et al*, 2021] algorithm has been applied to retrieve the right frequency configuration of each hyper-filtering layers. Finally, for each set of the so-processed hyper-filtered PPG-derived signals, we can compute for each waveform's sample *s(t$_k$)* a pattern-signal composed by the intensity-dynamic of that sample in each hy-per-filtered signal. We will thus obtain a large dataset of so built pattern-signals. The collected dataset of hyper-filtered pattern-signals will be classified by ad-hoc designed temporal-residual downstream deep classifier described in the next section.

## 2.2 The proposed 1D Deep Classifier

In Fig. 2 we have showed an overview of the proposed pipeline scheme including the designed Deep 1D Temporal Dilated Convolutional Neural Network (1D-CNN) suitable to classify the hyper-filtered signal patterns [Rundo *et al*, 2019b]. A temporal convolutional residual network embeds a dilated causal convolution layer capable of acting on the temporal stages of each set of wave sequences, has been implemented [Rundo *et al*, 2021; Rundo *et al*, 2019d]. The proposed 1D-CNN is composed of 12 residual blocks with a dilated convolution (3x3 kernel filters) followed by normalization, ReLU activations blocks and spatial dropout. The deep backbone includes a final softmax stage for data classification. For each of the blocks there is a progressive increase in the dilation starting from 2 and increases with a power of 2 till to 16. The output of the 1D-CNN is a binary assessment (0-0.5: Drowsy Driver ; 0.51-1: Wakeful Driver) of the driver's drowsiness level associated to the PPG signal of the monitored car driver. The described Deep Learning framework confirmed to be effective in assessing the driver's level of drowsiness as shown by the results reported in the following table.

| Method | Driver Attention Level Detection | |
|---|---|---|
| | *Drowsy Driver* | *Wakeful Driver* |
| **Proposed** | **98.71 %** | **99.03 %** |
| Shallow Network [Rundo et al, 2019b] | 96,50 % | 98,40 % |
| MLP | 92,22 % | 91,98 % |
| SVM | 90,11 % | 88,76 % |

Table 1: PPG-based Driver Drowsiness Monitoring Performance

As showed in Table 1, the implemented system outperformed similar approach based on classical machine learning methods such Support-Vector Machine (SVM), Multi-layer Perceptron (MLP), shallow-neural network [Rundo *et al*, 2021; Rundo *et al*, 2019d]. As introduced, the proposed car driver drowsiness monitoring has been combined with an intelligent driving scenario safety assessment based on pedestrians detection and tracking.

## 3. The *Criss-Cross* enhanced Mask-R-CNN for intelligent pedestrian tracking system

The following figure shows the proposed combined approach embedding the introduced intelligent PPG-based drowsiness monitoring system with enhanced Mask-R-CNN used for pedestrians' segmentation and tracking of the driving scenario.

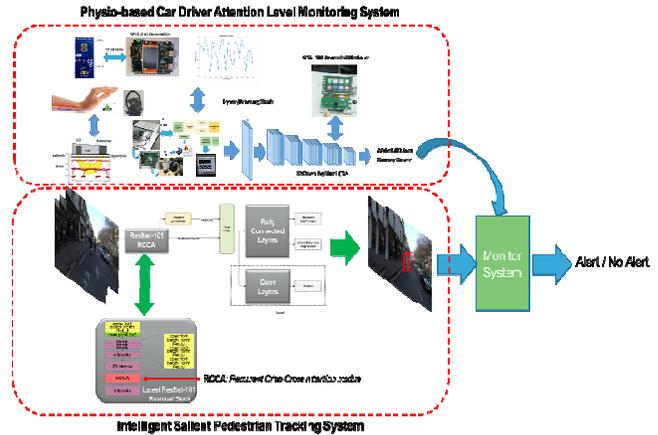

**Figure 3.** Overall scheme of the proposed pedestrians tracking pipeline

As schematized in Fig. 3, an enhanced Mask-R-CNN architecture [Rundo *et al*, 2021] embedding Self Attention is proposed. Mask-R-CNN is widely used in the automotive field. The target of the implemented enhanced Mask-R-CNN regards the ability to perform a pixel-based segmentation of the input image representing the driving scene. Moreover, with this solution we are able to generate the corresponding bounding-box that characterizes the Region of Interest (ROI) on which to perform post-processing. The features generator is based on classical DenseNet-201 [Rundo *et al*, 2021]. This deep classifier embeds a Recurrent Criss-Cross Attention (RCCA) layer [Rundo *et al*, 2021; Rundo *et al*, 2019d].

The attention mechanism based on Criss-Cross algorithm was first proposed in scientific literature [Rundo *et al*, 2021; Rundo *et al*, 2019d] showing very promising performance in several tasks including semantic segmentation. Specifically, the proposed Criss-Cross attention module is able to compute an innovative pixel-based contextual processing of the input image-frame.

As introduced, the enhanced Mask-R-CNN allows to obtain the bounding-box of the pedestrian needed to determine the distance from the driver's car. Quite simply, the height and width of the segmentation bounding box of each segmented pedestrian will be determined. Only bounding-boxes that have at least one of the two dimensions greater than two heuristically fixed thresholds ($L_1$ and $L_2$ respectively for length and width) will be considered *salient pedestrians*. Some instances in the next figures.

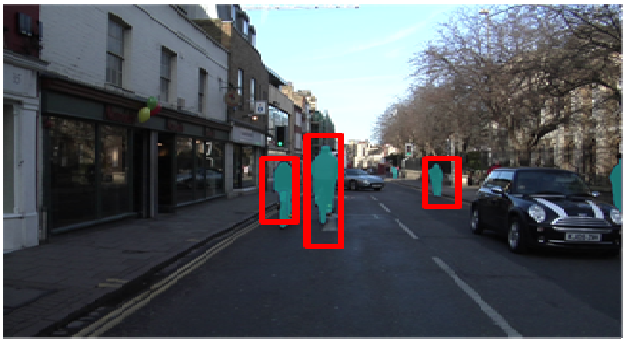
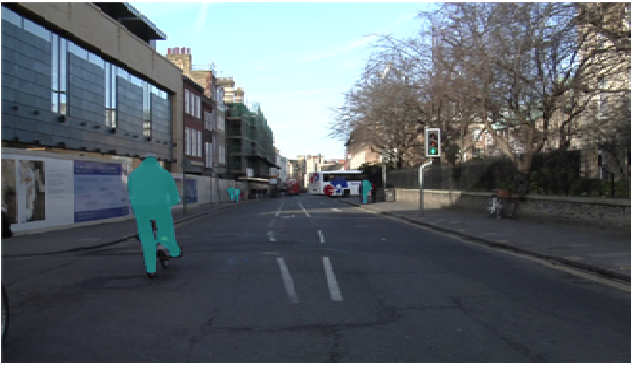

**Figure 4.**: Some instances of the driving scene frames overlayed with segmented salient pedestrian in different configurations (walking, by bicycle, etc ..). In red, the predicted bounding boxes segmentation.

As reported in the following, Table 2, the proposed deep system shows very interesting performance in CamVid dataset [Rundo *et al*, 2021]. Current research activity based on the usage of knowledge distillation algorithms is working over the proposed Mask-R-CNN in order to generate a compressed "student" model deployable over the mentioned automotive-grade MCUs delivered by STMicroelectronics.

| Method | Pedestrian Tracking System |
|---|---|
| | *mIoU* |
| **Proposed** | **69,70 %** |
| Faster-R-CNN (ResNet-50 backbone) | 53,95 % |
| Classical Fully Convolutional Network (ResNet-101 backbone) | 63,96 % |

Table 2: Experimental Results of the Pedestrian Tracking System

## 4. Conclusion and future works

The proposed system shows an interesting approach based on deep learning for monitoring the drowsiness level of the car driver. Through the outcomes of the physio-based car driver attention assessment combined with the designed driving scenario evaluation based on advanced AI-based computer vision algorithms, the author proposed an effective solution for the risk monitoring of the driving. The investigated deep architectures will be designed to be deployable to automotive-grade MCUs delivered by STMicroelectronics. The reported performance confirmed the effectiveness of the proposed approach. Future works aim to extend the drowsiness assessment of the car driver by integrating computer vision algorithms based on saliency analysis.

## 5. Acknowledgment - Funded Research project

This research was funded by the National Funded Program 2014-2020 under grant agreement n. 1733, (ADAS + Project).

## References

[1] [Tomokiko *et al*, 2015] Tomohiko Igasaki ; Kazuki Nagasawa ; Nobuki Murayama ; Zhencheng Hu. Drowsiness estimation under driving environment by heart rate variability and/or breathing rate variability with logistic regression analysis. In IEEE Proceedings of the International Conference on Biomedical Engineering and Informatics (BMEI), 2015;

[2] [Vicente *et al*, 2011] José Vicente ; Pablo Laguna ; Ariadna Bartra ; Raquel Bailón, Detection of driver's drowsiness by means of HRV analysis, in IEEE Proceedings of Computing in Cardiology, 2011;

[3] [Rundo *et al*, 2018] Rundo, F.; Conoci, S.; Ortis, A.; Battiato, S. An Advanced Bio-Inspired PhotoPlethysmoGraphy (PPG) and ECG Pattern Recognition System for Medical Assessment. Sensors 2018, 18, 405;

[4] [Vinciguerra *et al*, 2019] Vinciguerra, V., Ambra, E., Maddiona, L., Romeo, M., Mazzillo, M., Rundo, F., Fallica, G., di Pompeo, F., Chiarelli, A.M., Zappasodi, F., Merla, A., Busacca, A., Guarino, S., Parisi, A., Pernice, R., PPG/ECG multisite combo system based on SiPM technology, Lecture Notes in Electrical Engineering, 2019, Vol 539, pag. 105-109;

[5] [Mazzillo *et al*, 2018] Mazzillo, M., Maddiona, L., Rundo, F., Sciuto, A., Libertino, S., Lombardo, S., Fallica, G., Characterization of sipms with nir long-pass interferential and plastic filters, IEEE Photonics Journal, 2018, Vol. 10, Issue 3;

[6] [Conoci *et al*, 2018] Sabrina Conoci , Francesco Rundo , Giorgio Fallica , Davide Lena , Irene Buraioli , Danilo Demarchi, Live Demonstration of Portable Systems based on Silicon Sensors for the monitoring of Physiological Parameters of Driver Drowsiness and Pulse Wave Velocity, in IEEE Proceedings of Biomedical Circuits and Systems Conference (BioCAS), 2018;

[7] [Trenta *et al*, 2019a] F. Trenta, S. Conoci, F. Rundo and S. Battiato, "Advanced Motion-Tracking System with Multi-Layers Deep Learning Framework for Innovative Car-Driver Drowsiness Monitoring," 2019 14th IEEE International Conference on Automatic Face & Gesture Recognition (FG 2019), Lille, France, 2019, pp. 1-5, doi: 10.1109/FG.2019.8756566.

[8] [Rundo *et al*, 2019b] Rundo, F.; Spampinato, C.; Conoci, S. Ad-Hoc Shallow Neural Network to Learn Hyper Filtered PhotoPlethysmoGraphic (PPG) Signal for Efficient Car-Driver Drowsiness Monitoring. *Electronics* **2019**, *8*, 890.

[9] [Rundo *et al*, 2021] Rundo F, Conoci S, Spampinato C, Leotta R, Trenta F and Battiato S (2021) Deep Neuro-Vision Embedded Architecture for Safety Assessment in Perceptive Advanced Driver Assistance Systems: The Pedestrian Tracking System Use-Case. Front. Neuroinform. 15:667008. doi: 10.3389/fninf.2021.667008

[10] [Rundo *et al*, 2019c] Rundo, F.; Rinella, S.; Massimino, S.; Coco, M.; Fallica, G.; Parenti, R.; Conoci, S.; Perciavalle, V. An Innovative Deep Learning Algorithm for Drowsiness Detection from EEG Signal. *Computation* 2019, 7, 13.

[11] [Rundo *et al*, 2019d] Rundo, F., Petralia, S., Fallica, G., Conoci, S., A nonlinear pattern recognition pipeline for PPG/ECG medical assessments, Lecture Notes in Electrical Engineering, 2019, Vol 539, pag. 473-480;